\documentclass{article}

\usepackage[pdftex]{graphicx}
\graphicspath{}
\DeclareGraphicsExtensions{.pdf}

\usepackage{amsmath}
\usepackage[caption=false,font=footnotesize]{subfig}
\usepackage{acronym}
\usepackage{authblk}

\DeclareMathOperator\erf{erf}

\begin{document}

\title{Potholes Ahead: Impact of Transient \\ Link Blockage on Beam Steering in \\ Practical mm-Wave Systems}

\author[1]{Adrian Loch\thanks{adrian.loch@imdea.org}}
\author[1, 2]{Irene Tejado\thanks{irene.tejado@imdea.org}}
\author[1]{Joerg Widmer\thanks{joerg.widmer@imdea.org}}
\affil[1]{IMDEA Networks Institute, Madrid, Spain}
\affil[2]{Universidad Carlos III, Madrid, Spain}

\renewcommand\Authands{ and }
\date{}

\maketitle

\acrodef{snr}[SNR]{Signal-to-Noise Ratio}
\acrodef{mcs}[MCS]{Modulation and Coding Scheme}
\acrodef{tcp}[TCP]{Transmission Control Protocol}
\acrodef{cots}[COTS]{commercial off-the-shelf}
\acrodef{bbs}[BBS]{Blind Beam Steering}
\acrodef{ws}[WS]{window size}
\acrodef{mmwave}[mm-Wave]{millimeter wave}
\acrodef{usrp}[USRP]{Universal Software Radio Peripheral}
\acrodef{bi}[BI]{Beacon Interval}
\acrodef{ssi}[SSI]{Sector Sweep Interval}
\acrodef{csmaca}[CSMA/CA]{Carrier Sense Multiple Access with Collision Avoidance}

\begin{abstract}
The practical realization of beam steering mechanisms in millimeter wave communications has a large impact on performance. The key challenge is to find a pragmatic trade-off between throughput performance and the overhead of periodic beam sweeping required to improve link quality in case of transient link blockage. This is particularly critical in commercial off-the-shelf devices, which require simple yet efficient solutions. First, we analyze the operation of such a commercial device to understand the impact of link blockage in practice. To this end, we measure TCP throughput for different traffic loads while blocking the link at regular intervals. Second, we derive a Markov model based on our practical insights to compute throughput for the case of transient blockage. We use this model to evaluate the trade-off between throughput and periodic beam sweeping. Finally, we validate our results using throughput traces collected using the aforementioned commercial device. Both our model and our practical measurements show that transient blockage causes significant signal fluctuation due to suboptimal beam realignment. In particular, fluctuations increase with traffic load, limiting the achievable throughput. We show that choosing lower traffic loads allows us to reduce fluctuations by 41\% while achieving the same net throughput than with higher traffic loads.
\end{abstract}


\section{Introduction}
\label{sec:intro}

Distinguishing link degradation due to mobility from that due to blockage in \ac{mmwave} networks is challenging. However, transceivers operating in this band must be able to tell both apart for proper beam steering. Due to the high path loss at \ac{mmwave} frequencies, transceivers use beamforming to overcome attenuation \cite{challenges60Ghz}. Both the transmitter and the receiver must steer their beams towards each other to achieve a high \ac{snr}. If one of them moves, they must adjust their beams accordingly. However, in case of transient link blockage due to, e.g., a person crossing the link, both transceivers should maintain their original beam steering. In home or office scenarios, such as envisioned in 802.11ad \cite{ad_standard}, mobility and blockage are likely to happen often.

\begin{figure}
	\centering
		\includegraphics{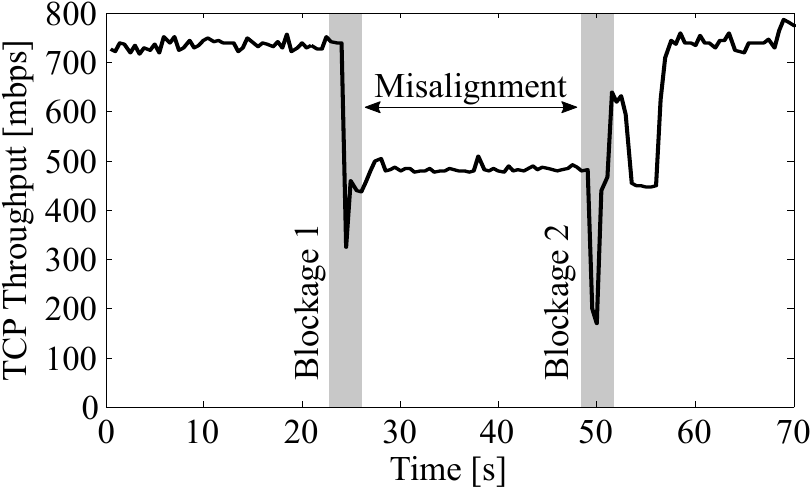}		
	\caption{Throughput ``pothole'' during subsequent transient blockage events.}
	\label{fig:example_blockage}	
\end{figure}

Figure~\ref{fig:example_blockage} shows an example of the impact of transient blockage on the \ac{tcp} throughput for a \ac{cots} 60 GHz device. While related work also studies the impact of blockage on 60 GHz communications \cite{Zhu2014Mobicom, measurements_blockage}, it does not consider the beam steering misalignment that may occur as a result. In Figure~\ref{fig:example_blockage}, we observe that after the first blockage at second $25$, the throughput stabilizes at more than $200$ mbps less than that prior to the blockage, even though the link is unobstructed. In other words, our \ac{cots} device has interpreted the first blockage as mobility, thus trying to adapt its beam steering. While the new steering allows for communication, the \ac{snr} is lower, resulting in lower \ac{tcp} throughput. At second 50 in Figure~\ref{fig:example_blockage}, a second blockage causes another beam realignment, which in this case fortunately results in the original beam steering and allows \ac{tcp} to achieve again roughly $700$ mbps.
Figure~\ref{fig:example_blockage} shows that the impact of transient blockage can be highly detrimental to throughput if the transceiver classifies blockage as mobility, causing a throughput ``pothole'' while the link is misaligned. First, the link operates significantly below its potential. Second, the throughput exhibits high fluctuations. Regarding the latter, we find that throughput can stabilize at a number of different levels after a blockage. This is critical for traffic which requires stable, high throughput links, such as uncompressed video.


Related work proposes a number of beam steering solutions that partially solve the above problem. The 802.11ad standard suggests beam tracking to address mobility. That is, devices continuously track \ac{snr} variations and adapt their beam steering accordingly. This avoids costly beam sweeping, i.e., exhaustively probing all directional beam patterns of a transmitter to find the receiver after the transmitter and/or the receiver move. Still, beam tracking itself cannot distinguish mobility from blockage, resulting in the behavior in Figure~\ref{fig:example_blockage}. Similarly, other approaches that aim at reducing the complexity of beam sweeping \cite{opt_beam_sweep} cannot distinguish them neither. In contrast, \ac{bbs} \cite{bbs} estimates the direction in which the receiver is located. Essentially, the transmitter uses an antenna array operating at 2.4 GHz to estimate the angle-of-arrival, and then uses this information to perform beam steering at 60 GHz. Hence, transceivers can distinguish mobility from blockage. However, \ac{bbs} considers a 2.4 GHz antenna array with at least four antennas, which at a minimum antenna separation of $\lambda/2$ might be impractical for certain devices. Other approaches exploit information about transmit directions which worked prior to an \ac{snr} drop to narrow down the exhaustive beam steering search \cite{mobiwac}. Alternatively, geometric analysis can also reveal which alternative beams are available at any location within a room \cite{beamscope}. However, such approaches are meaningful for long-lived blockages which require finding alternative non-line-of-sight paths, while we consider a \emph{transient} blockage scenario. Recent work \cite{wimi} also suggests analyzing the initial samples of an \ac{snr} drop in order to determine whether it is due to mobility or blockage. Such an approach must operate at the physical layer since it requires a timely reaction, i.e., the device must identify the cause of the \ac{snr} drop immediately when it starts. This strict time constraint  may hinder its implementation in \ac{cots} devices. Although the propagation characteristics in the 60 GHz band allow for precise tracking \cite{mtrack}, distinguishing mobility from blockage remains challenging.

As a result of the above challenge, \ac{cots} devices typically implement simple heuristics that result in the fluctuating behavior shown in Figure~\ref{fig:example_blockage}. In this paper, we analyze such throughput fluctuations and show their relation to frame-level aggregation \cite{aggregation_overview}. Specifically, our contributions are as follows:

\begin{itemize}

	\item We measure \ac{tcp} throughput, \ac{mcs} adaptation, frame-level aggregation, and beam steering for a \ac{cots} device to understand the impact of transient link blockage in practice.
	\item We derive an analytical Markov chain model for transient link blockage based on our practical insights. We use it to analyze the trade-off between throughput and the overhead caused by periodic beam sweeps.	
	\item We show that frame aggregation can compensate for \ac{mcs} drops due to steering misalignments, thus reducing throughput fluctuations by up to $41\%$ while maintaining high throughput on \ac{cots} 60 GHz devices.
\end{itemize}

The remainder of this paper is structured as follows. In Section~\ref{sec:problem} we analyze the practical observations that allow us to understand the impact of transient blockages. After that, Section~\ref{sec:approach} presents our Markov chain model, and Section~\ref{sec:results} discusses the results that we obtain both from the model and from a \ac{cots} device. Finally, Section~\ref{sec:conc} concludes the paper.


\section{Practical Observations}
\label{sec:problem}

In this section, we use a 60 GHz \ac{cots} device to understand the impact of transient link blockage. To this end, we set up a 60 GHz link and walk through it at a normal indoor pace. We perform the experiment for low and high link loads.

\subsection{Experiment setup}
\label{subsec:observations}

\ac{cots} networking devices that operate at \ac{mmwave} frequencies are not yet widely available. However, some notebook manufacturers have implemented the WiGig \cite{wigig_overview} standard---which is very similar to 802.11ad---to design wireless docking stations. Basically, docking station and notebook establish a 60 GHz link to replace the traditional physical connection between both devices. We use such a setup for our practical experiments. In particular, we place a Dell D5000 docking station and a Dell Latitude E7440 notebook on two tables and separate them about two meters. We use iperf to transmit data on the 60 GHz link connecting both devices. To control the traffic load, we adjust the \ac{tcp} \ac{ws}. For each experiment, we record the signal level, the \ac{mcs}, the \ac{tcp} throughput, and the frame aggregation size. To record the signal level, we overhear the communication using a SiversIMA FC1005V/00 V-band converter connected to a \ac{usrp} X310. While the \ac{usrp}'s bandwidth does not allow us to decode data, we obtain energy level traces that allow us to infer the beam alignment and the frame aggregation size.

\subsection{Effects at the application layer}
\label{subsec:appl_layer_effects}

\begin{figure}
	\centering
		\includegraphics{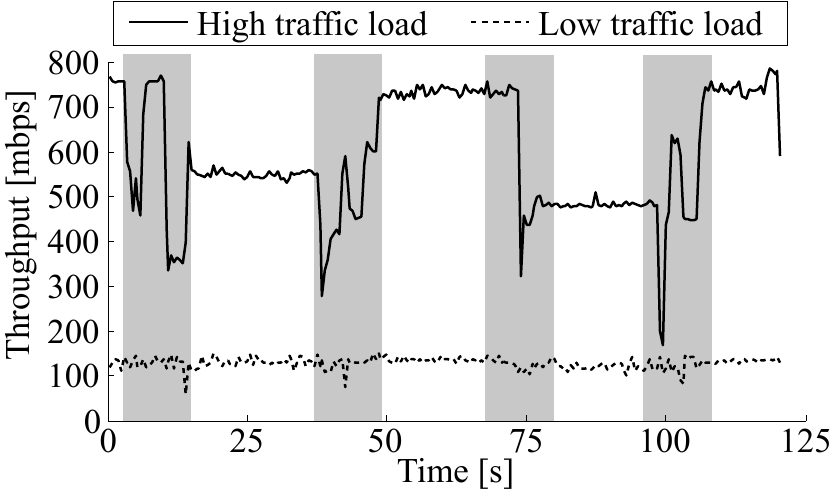}
	\caption{Suboptimal throughput levels after blockages.}
	\label{fig:example_trace}
\end{figure}

Figure~\ref{fig:example_trace} shows the application layer TCP throughput for both a high and a low link load. In this case, we measure the throughput every $500$ ms, and induce transient link blockages roughly every $20$ to $30$ seconds. Each time we cross the link, we observe a significant throughput drop. Moreover, for increasing link loads our measurements show that the throughput drops at each blockage become larger and the number of suboptimal levels increases. As sketched in Section~\ref{sec:intro}, the highest level corresponds to the best beam alignment for a certain load. The other levels correspond to suboptimal beam alignments resulting in a range of \ac{mcs} levels. While Figure~\ref{fig:example_trace} clearly shows that transient link blockages have a larger impact at higher traffic loads, the underlying reason is not evident. In the following, we show that this effect is related to aggregation.

\subsection{Analysis at the lower layers}

\begin{figure*}[!t]
\centerline{\subfloat[Signal level]{\includegraphics{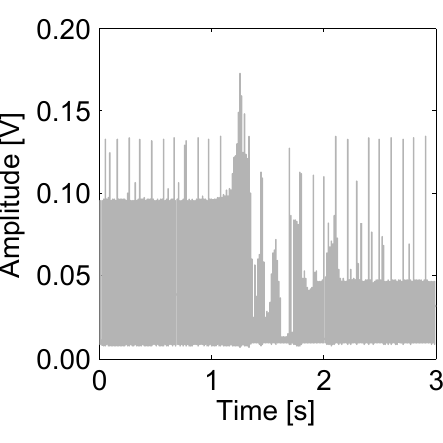}
\label{fig:block_w10_trace}}
\hfil
\subfloat[Modulation/coding scheme]{\includegraphics{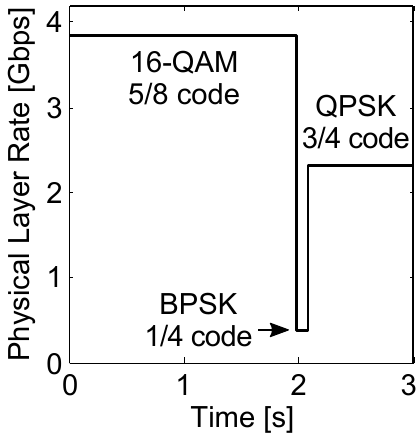}
\label{fig:block_w10_mcs}}
\hfil
\subfloat[Iperf throughput]{\includegraphics{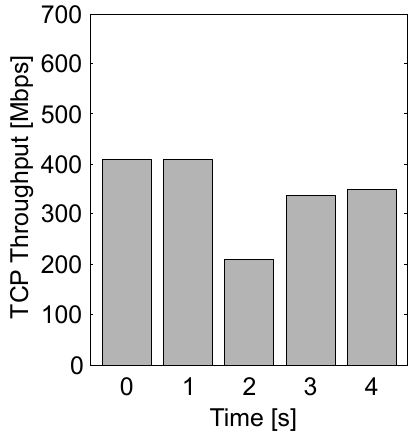}
\label{fig:block_w10_iperf}}
\hfil
\subfloat[Frame aggregation]{\includegraphics{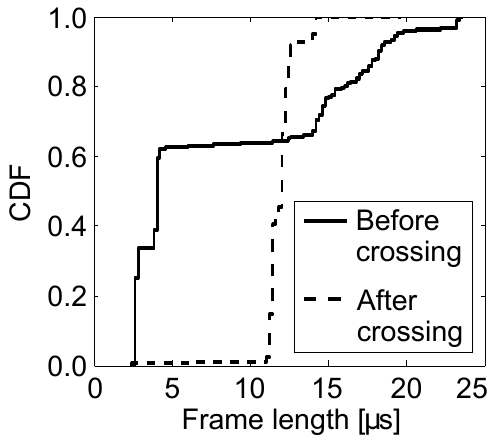}
\label{fig:block_w10_aggr}}}
\caption{Blockage with low link load in practice. The long blockage in (a) is likely due to the automatic gain control adjustment of the SiversIMA converter.}
\label{fig:block_w10}
\vspace{-2mm}
\end{figure*}

\begin{figure*}[!t]
\centerline{\subfloat[Signal level]{\includegraphics{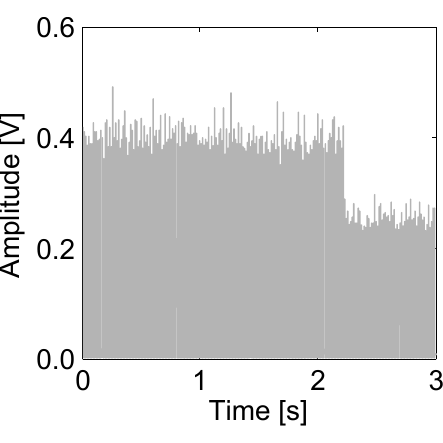}
\label{fig:block_w100_trace}}
\hfil
\subfloat[Modulation/coding scheme]{\includegraphics{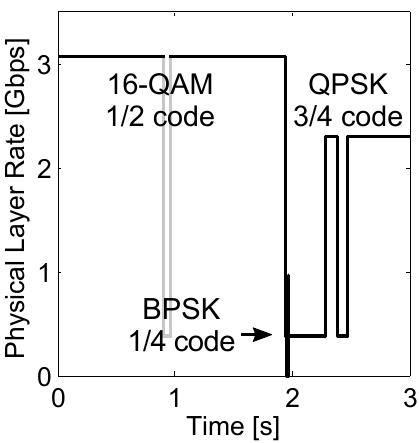}
\label{fig:block_w100_mcs}}
\hfil
\subfloat[Iperf throughput]{\includegraphics{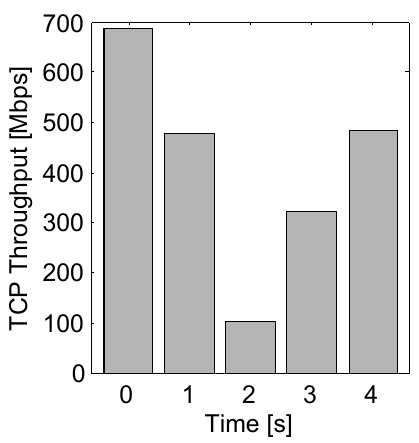}
\label{fig:block_w100_iperf}}
\hfil
\subfloat[Frame aggregation]{\includegraphics{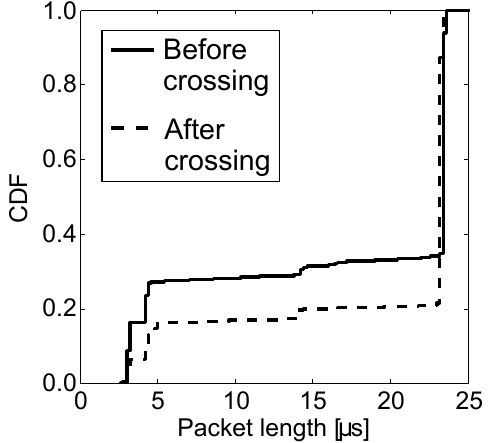}
\label{fig:block_w100_aggr}}}
\caption{Blockage with high link load in practice. The brief \ac{mcs} drop grayed out in (b) is not due to the blockage but occurred by chance in this measurement.}
\label{fig:block_w100}
\vspace{-1mm}
\end{figure*}

To understand the reason for the behavior in Figure~\ref{fig:example_trace}, we measure lower layer metrics during a transient link blockage. Figure~\ref{fig:block_w10} and Figure~\ref{fig:block_w100} show these metrics for the case of low (\ac{ws} 10 kB) and high (\ac{ws} 100 kB) link load, respectively.

\subsubsection{Low link load}

At low load, the link achieves a rate of roughly $400$ Mbps before the blockage, as shown in Figure~\ref{fig:block_w10_iperf}. As soon as we walk across the link, the signal level shown in Figure~\ref{fig:block_w10_trace} drops significantly. However, after the blockage, it only rises again to about half of the amplitude. This indicates that the docking station has switched to a different sector due to the blockage. Since the docking station is steering in a different direction while the transmit power remains unchanged, we observe a different signal level at the SiversIMA converter. In other words, the signal level drop in Figure~\ref{fig:block_w10_trace} has triggered a sector level sweep which has led to a suboptimal sector since the person was still crossing the link when it took place. As a result, the \ac{snr} at the receiver dropped and thus the link switched to a significantly lower \ac{mcs}, as shown in Figure~\ref{fig:block_w10_mcs}. Still, while the physical layer rate has nearly halved, the iperf throughput in Figure~\ref{fig:block_w10_iperf} stabilizes at only $14.5\%$ less throughput than prior to the blockage. 

Figure~\ref{fig:block_w10_aggr} explains this somewhat surprising behavior. Before the blockage occurs, the duration of more than $60\%$ of all data frames is less than $5\,\mu s$. However, after the blockage most frames are longer than $10\,\mu s$. That is, frame aggregation increases. The Dell D5000 is known to simply aggregate all data packets available in the transmit queue whenever it gets access to the channel. Due to the drop in \ac{mcs} after blockage, the transmission time increases. During the transmission, packets continue arriving at the queue at the same pace. Hence, the average queue length after the blockage is larger than before the blockage, resulting in larger aggregated frames. This means that sector mismatches due to transient link blockages have a small impact on application layer throughput for low link loads because the frame aggregation capability of the 802.11ad standard compensates for it.

\subsubsection{High link load}
\label{subsubsec:obs_high_load}

For high link load, in Figure~\ref{fig:block_w100_trace} we observe again a sector mismatch after the blockage. As expected, also the \ac{mcs} in Figure~\ref{fig:block_w100_mcs} drops. Still, in Figure~\ref{fig:block_w100_iperf} we observe that the impact on application layer throughput is much larger than for the low link load case---the throughput drops by roughly 30\%. The underlying reason is that, at high link load, the docking station is already using a high level of aggregation. As shown in Figure~\ref{fig:block_w100_aggr}, approximately $70\%$ of all data frames are longer than $20\mu s$ before the blockage. Thus, the D5000 can only increase aggregation by about $10\%$, which is not enough to compensate for the lower \ac{mcs}. This suggests that increasing the traffic load beyond a certain level might not pay off for a given frequency of transient link blockages.


\section{System Model}
\label{sec:approach}

From Section~\ref{sec:problem} we conclude that transient blockage may cause sector misalignment which in turn affects \ac{mcs} and ultimately frame aggregation. Aggregation can mitigate blockage at moderate link loads. This is not intrinsic to the D5000 since any \ac{mmwave} device must deal with the above issues. In the following, we derive a Markov model based on these insights.

\subsection{Model overview}
\label{sec:model_ovw}

\begin{figure}
	\centering
		\includegraphics{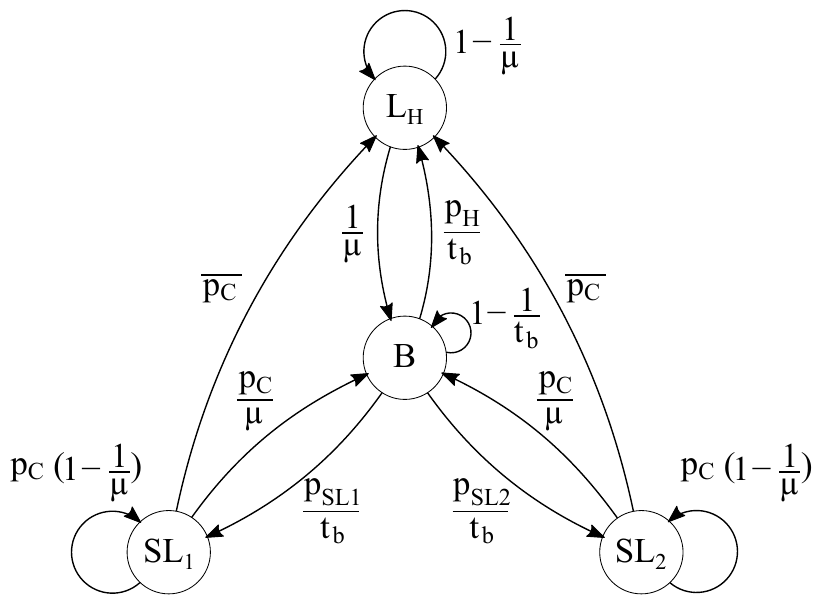}
	\caption{Example Markov model for a link experiencing transient blockages.}
	\label{fig:example_markov}
\end{figure}

Due to suboptimal sector alignments, a link may be in $N+1$ different states: $N$ states due to all feasible combinations of transmit and receive sectors, and one state when the link is blocked. Figure~\ref{fig:example_markov} shows an example of such a Markov model for a link with $N = 3$ throughput levels, that is, one high level $\text{L}_\text{H}$, and two suboptimal levels $\text{SL}_\text{1}$ and $\text{SL}_\text{2}$. The blockage state is named $\text{B}$. Whenever a transient link blockage occurs, the model transitions to blockage state $\text{B}$. From there, it moves to one of the $N$ states until the link is blocked again. If the state is one of the suboptimal ones, the model directly transitions to $\text{L}_\text{H}$ whenever the device performs a periodic sector level sweep. We choose a time-slot size of one millisecond for our Markov model since this matches the timescale at which  sector level sweeps are expected to occur. However, it is straightforward to adjust the model parameters for other time-slot sizes.

We model the occurrence of transient link blockage as a Gaussian distribution $\mathcal{N}(\mu, \sigma^2)$. The probability $p_S$ of staying in state $\text{L}_\text{H}$ for one time-slot is directly related to $\mu$. To derive it, we first formulate the probability $P(k)$ of staying $k > 1$ time-slots at that state as in Equation~\ref{equ:pk_ps}. Then, we compute the average time $T$ that we stay at the state as in Equation~\ref{equ:avg_time_state}.
\begin{equation}
	P(k) = p_S^{k-1} (1-p_S)
\label{equ:pk_ps}
\end{equation}
\begin{equation}
	T = \displaystyle\sum_{k=1}^{\infty} k P(k) = \displaystyle\sum_{k=1}^{\infty} k \cdot p_S^{k-1} \cdot (1-p_S) = \frac{1}{1-p_S}
\label{equ:avg_time_state}
\end{equation}
Since blockage occurs on average every $\mu$ seconds, $T = \mu$ and thus the probability of staying at $\text{L}_\text{H}$ is $p_S = 1 - 1/\mu$. Note that the units of $\mu$ are time-slots, i.e., milliseconds in our case. From this we directly obtain that the transition probability from $\text{L}_\text{H}$ to $\text{B}$ is $1 - p_S = 1/\mu$. We follow a similar reasoning to compute the probabilities of staying at state $\text{B}$. If blockages last on average $t_b$, the probability of staying at state $\text{B}$ is $1 - 1/t_b$. However, after a blockage, the link may recover to any of the other $N$ states. Hence, we weight the probability of leaving state $\text{B}$, i.e., $1/t_b$, with the probabilities of transitioning to each of the states. In the example in Figure~\ref{fig:example_markov}, $p_\text{H}$ is the probability to transition to $\text{L}_\text{H}$ while $p_{\text{SL1}}$ and $p_{\text{SL2}}$ are the probabilities of transitioning to each of the suboptimal states, respectively. In general, $p_\text{H} + \sum_{\forall i} p_{\text{SL}i} = 1$ must hold.


Finally, the probability of staying at the suboptimal states is related to $\mu$ in the same manner as for $\text{L}_\text{H}$. Still, in this case we must take into account that the device may perform a periodic sector level sweep that triggers a transition to $\text{L}_\text{H}$. We define $p_C$ as the probability that a transient link blockage occurs before the next periodic sector level sweep takes place. Hence, the transition probability from $\text{SL}_i$ to $\text{L}_\text{H}$ is $\overline{p_C}$. Since the probability of a sector sweep and a blockage are independent, we weight the result of Equation~\ref{equ:avg_time_state} with $p_C$ to obtain the probabilities of leaving and staying at a suboptimal state.


\subsection{Periodic sector level sweeps}

In the following, we derive the analytic expression of $p_C$ based on the statistical characteristics of blockages and the frequency of sector level sweeps. We consider that devices perform such sector level sweeps at regular intervals of length $S$. For 802.11ad, $S$ translates directly into the \ac{bi}, that is, the interval at which a station may transmit beacons to improve beam steering. Without loss of generality, we define the time origin $t = 0$ as the point in time when the last sector level sweep took place. Given that a blockage occurs at time $t = t_b$, the time that remains until the next sector level sweep is $S - t_b$. Hence, the probability of another blockage $C$ occurring prior to the next sector level sweep for time $t_b$ is as in Equation~\ref{equ:prob_cross_before_sweep}, where $\erf$ is the error function.

\vspace{-4mm}
\begin{equation}	
	\begin{split}
		P(C, t = t_b)	& = \int_0^{S-t_b} \frac{1}{\sigma \sqrt{2\pi}} e^{-\frac{(t-\mu)^2}{2\sigma^2}} \, \mathrm{d}t \\[1.5mm]
					& = \frac{1}{2} \left( \erf{\left( \frac{\mu}{\sqrt{2}\sigma} \right)} - \erf{\left( \frac{\mu-S+t_b}{\sqrt{2}\sigma} \right)} \right)
	\end{split}	
\label{equ:prob_cross_before_sweep}
\end{equation}
\vspace{-2mm}

Next, we obtain the probability in Equation~\ref{equ:prob_cross_before_sweep} for any value of $t_b \in \{0, S\}$. 
Using the definition of conditional probability, we obtain $P(C, t_b) = P(C, t = t_b) P(t_b)$, where $P(t)$ is evenly distributed because sector level sweeps occur at fixed intervals. 
%
%
Hence, $p_C$ is the marginal probability of $P(C, t_b)$ for all values of $t_b$. Based on this, we compute $p_C$ as in Equation~\ref{equ:final_pc_int}.

\vspace{-4mm}
\begin{equation}
\begin{aligned}	
		p_C 	= &\int_0^{S} P(C, t_b)\, \mathrm{d}t_b = \int_0^{S} P(C, t = t_b) \cdot P(t_b)\, \mathrm{d}t_b \\[1mm]
		   	= &\int_0^{S} \frac{1}{2} \left( \erf{\left( \frac{\mu}{\sqrt{2}\sigma} \right)} - \erf{\left( \frac{\mu-S+t_b}{\sqrt{2}\sigma} \right)} \right) \cdot \frac{1}{S}\, \mathrm{d}t_b \\[3mm]
		   	= &\left( \frac{S-\mu}{2S} \right) \left( \erf{\left( \frac{\mu}{\sqrt{2}\sigma} \right)} - \erf{\left( \frac{\mu - S}{\sqrt{2}\sigma} \right)} \right) - \\[1mm]
		   	& \sqrt{\frac{2}{\pi}} \frac{\sigma}{2S} \left( e^{-\frac{\mu^2}{2\sigma^2}} - e^{-\frac{\left(\mu - S\right)^2}{2\sigma^2}} \right)
\label{equ:final_pc_int}
\end{aligned}
\end{equation}
\vspace{-4mm}


\subsection{Throughput}

Next, we derive the throughput for each of the $N+1$ states of our Markov model. We consider a 60 GHz transmitter that operates as observed in Section~\ref{sec:problem}. That is, the transmitter has a queue of size $q$. Each time it gets access to the channel, the device aggregates up to $\text{a}_\text{max}$ data from the queue and transmits it. While it is transmitting, more packets arrive at the queue. Hence, the size of the aggregated frames directly depends on the transmission time $\text{t}_\text{TX}$ and the channel access time $\text{t}_\text{acc}$. The latter includes the \ac{csmaca} overhead as in 802.11ad. Further, control messages increase each \ac{bi} by a factor $f$.

The application generates a link load $l$. The maximum link load that the device supports occurs when it aggregates $\text{a}_\text{max}$ data in each transmission. We compute $l$ for a certain aggregation factor $\text{a}_\text{factor} \in \{0, 1\}$ such that the device aggregates $\text{a}_\text{max}$ for $\text{a}_\text{factor} = 1$. In this case, blockage has a significant impact since the device cannot aggregate more to mitigate the impact of lower \ac{mcs} values (c.f. Section~\ref{sec:problem}). Based on $l$, we compute the queue level $q_L$ for each of the $N$ states. Each state is related to a certain \ac{mcs}. For transmission $i$, the queue level is the transmission time $\text{t}_\text{TX}$ of the previous transmission multiplied by the link load $l$, as shown in Equation~\ref{equ:queue_levels}.

\begin{equation}
	q_{L,i} = \text{t}_{\text{TX}, i} \cdot l = \left( \frac{q_{L,i-1}}{\text{MCS}} + \text{t}_\text{acc} \right) \cdot f \cdot l
\label{equ:queue_levels}
\end{equation}

While $l \leq \text{MCS}$, the above equation converges to a stable queue level $q_L$ for $i \rightarrow \infty$. Based on this, we derive the throughput $\text{THP}$ for each state. If the queue level is less than or equal to $\text{a}_\text{max}$, the throughput is directly the link load $l$, as shown in Equation~\ref{equ:thp_stable}. However, if $q_L > \text{a}_\text{max}$, the application is generating more data than the link can transmit. Eventually, the queue level reaches its maximum length $q$. In this case, we limit the amount of data in each transmission to $\text{a}_\text{max}$.

\vspace{-3mm}
\begin{equation}
\text{THP}_\text{N} = \frac{q_L}{\text{t}_\text{TX}} =
  \begin{cases}
    \dfrac{\text{t}_\text{TX} \cdot l}{\text{t}_\text{TX}} = l       & \quad \text{if } q_L \leq \text{a}_\text{max}\\[4mm]
    \dfrac{\text{a}_\text{max}}{\left( \frac{\text{a}_\text{max}}{\text{MCS}} + \text{t}_\text{acc} \right) f}  & \quad \text{if } q_L > \text{a}_\text{max}\\
  \end{cases}
  \label{equ:thp_stable}
\end{equation}

Finally, we compute the throughput in state $\text{B}$. To this end, we define the blockage duration $t_B$ as the time during which the blockage affects the throughput of the link. In general, $t_B$ is the sum of (a) the time $t$ that the link is physically blocked and (b) the time to transmit the data that accumulates at the application while the link is blocked. The latter includes the new data $\text{d}_\text{new}$ that arrives while serving the accumulated data. Further, the accumulated data is limited to the queue size $q$. If the new state to which the Markov model transitions after a blockage already operates at the maximum aggregation size, (b) is infinite since the link cannot transmit more data. In that case, the duration of the blockage effect is just (a) and the related throughput is zero. Otherwise, we observe a higher throughput while (b) is ongoing. In this case, we compute the throughput in state $\text{B}$ as in Equation~\ref{equ:thp_blocked}. Since the duration of (b) depends on the \ac{mcs} of the state to which the Markov model transitions after the blockage, we compute $t_B$ as the average of all $t_{B_i}$ values for $j \in [1, N]$ weighted with the probabilities $p_j \in \{p_\text{H}, p_{\text{SL}i}\}$ of transitioning to each of the states.


\vspace{-0.7mm}
\begin{equation}	
	\text{THP}_\text{B} = \displaystyle\sum_{\forall j} p_j \frac{\min(t \cdot l, q) + \text{d}_\text{new}}{t_{B_j}}
\label{equ:thp_blocked}
\end{equation}


\section{Evaluation}
\label{sec:results}

Using our Markov model in Section~\ref{sec:approach}, we study the impact of transient link blockage both in theory and based on practical traces from our 60 GHz wireless docking station.


\subsection{Analytical evaluation}

For our analytical evaluation, we set the parameters of our model as in Table~\ref{tab:params_anlyt}. We consider a 60 GHz link with $N = 3$ states. Whenever a blockage occurs, the probability of transitioning to each of the states is the same. We evaluate the impact of four parameters, namely, the \ac{ssi}, $\mu$, $\text{a}_\text{factor}$, and $v$. Parameter $v$ is the speed at which a person causing the blockage walks across the link. We use it to compute $t$, i.e., the amount of time that the link is physically blocked. To this end, we assume a beamwidth of $\alpha$ and that the person crosses the link at a distance $d$ of the transmitter.

\begin{table}
\renewcommand{\arraystretch}{1.3}
\caption{Evaluation Parameters (Defaults Underlined)}
\label{table_example}
\centering
\begin{tabular}{l|l||l|l}
\hline
\bfseries Parameter & \bfseries Value & \bfseries Parameter & \bfseries Value  \\
\hline\hline
$\text{t}_\text{DIFS}$ 	& $13 \mu$s 					& $\alpha$ 				& $20^{\circ}$\\
$\text{t}_\text{SIFS}$ 	& $3 \mu$s 					& $d$ 					& $1.5$\\
$\text{t}_\text{ACK}$ 		& $6 \mu$s 					& $v$ 					& $[0.1\text{m/s}, 4\text{m/s}]$ Default: \underline{$1\text{m/s}$}\\
$\text{t}_\text{SLOT}$ 	& $5 \mu$s 					& MCS 					& $3.85 \text{Gbps}, 1.925 \text{Gbps}, 1.155 \text{Gbps}$\\
$\text{CW}_\text{min}$ 	& $15$ 						& $\text{p}_\text{H}$ 		& $1/3$\\	
$\text{t}_\text{sweep}$	& $4\text{ms}$ 				& $\text{p}_\text{SL1}$,  $\text{p}_\text{SL2}$ 	& $1/3$, $1/3$\\
SSI 						& $[10\text{ms}, $\underline{$\infty$}$]$  		& $q$ 	& $793.5\text{kB}$\\
$\mu$ 					& $[2 \text{s}, 20 \text{s}]$ 	& $\text{a}_\text{factor}$ 	& $0.2, 0.4, 0.6, 0.8, $ \underline{$1.0$}\\
$\sigma$ 					& $0.1 \text{s}$ 				& $\text{a}_\text{max}$ 	& $79.35\text{kB}$ \\
\hline
\end{tabular}
\label{tab:params_anlyt}
\end{table}

\subsubsection{Link load} 

\begin{figure}
	\centering
		\includegraphics{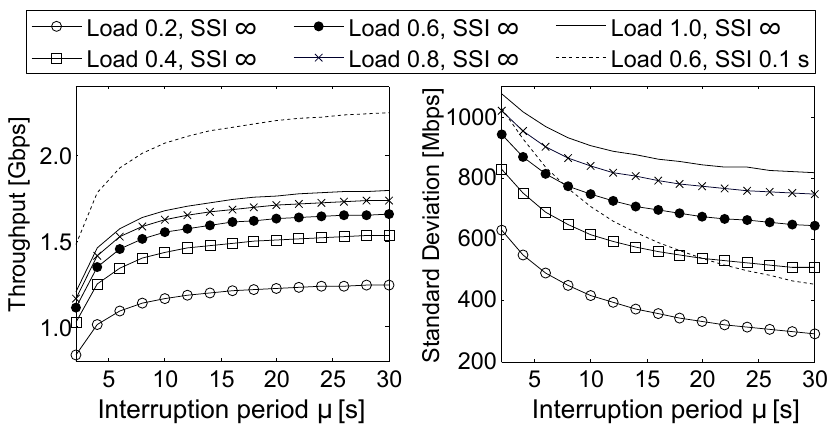}
	\caption{Throughput and throughput fluctuation for different link traffic loads.}
	\label{fig:anlyt_load}
\end{figure}

Figure~\ref{fig:anlyt_load} depicts our results for increasing link loads and average blockage intervals $\mu$. As a starting point, we set the \ac{ssi} to infinity, that is, the device only switches to a different sector when a blockage occurs. While the average throughput increases with the link load, the improvement becomes very small as the load approaches one. As discussed in Section~\ref{sec:problem}, this occurs because aggregation cannot mitigate the impact of lower \ac{mcs} values. Further, Figure~\ref{fig:anlyt_load} also shows that the throughput fluctuation increases significantly with the link load. For instance, while increasing the link load from $0.4$ to $0.6$ improves throughput by just $8\%$ for $\mu = 20$, it worsens throughput fluctuations by $25\%$. Hence, increasing the link load would hardly pay off in this case. For comparison, we include the result for an \ac{ssi} of $100$ ms. As expected, the average throughput is much higher since the link quickly switches back to the best sector after a blockage. Also, the variance decreases much faster with $\mu$ because the transmitter is using the same sector most of the time.

\subsubsection{\acl{ssi}}

\begin{figure}[t]
	\centering
		\includegraphics{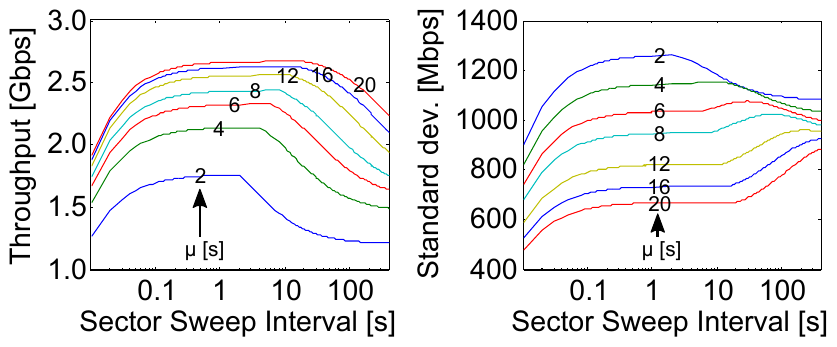}
	\caption{Throughput and throughput fluctuation for different \ac{ssi} and $\mu$ values.}
	\label{fig:anlyt_dti}
\end{figure}

In our second analysis, we study the impact of the \ac{ssi}. Figure~\ref{fig:anlyt_dti} depicts a trade-off regarding how often a device performs sector level sweeps. If the \ac{ssi} is very short, the resulting overhead limits the throughput significantly. Conversely, if it is too long, the blockages also reduce the average throughput. We achieve the best performance when the \ac{ssi} matches the average blockage interval $\mu$. Hence, a meaningful strategy for 60 GHz devices would be to estimate how often significant \ac{snr} drops occur in a given scenario, and set the \ac{ssi} to that value. Regarding throughput fluctuations, very short \ac{ssi} values are best since this ensures that the link is most of the time in the best state. As the \ac{ssi} increases, the variance becomes larger because the link is more often in suboptimal states. Beyond $\text{\ac{ssi}} = \mu$, the variance increases again for large $\mu$ since link state changes are mostly due to blockages. While sector sweeps ensure that the link recovers to the best state, blockages may lead to suboptimal states. 

\subsubsection{Speed}

\begin{figure}[t]
	\centering
		\includegraphics{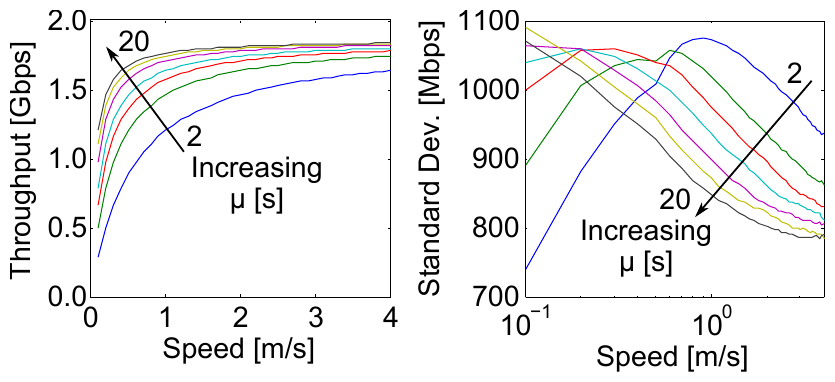}
	\caption{Throughput and throughput fluctuation for different $v$ and $\mu$ values.}
	\label{fig:anlyt_speed}
\end{figure}

Next, we study how the speed of the person causing the blockage affects throughput. In Figure~\ref{fig:anlyt_speed}, we observe that the higher the speed, the higher the throughput is. This is expected, since the person blocks the link for a shorter amount of time $t$. Regarding variance, fluctuations generally decrease with speed because the system stays less time in state $\text{B}$, which often only achieves very low or even zero throughput. Initially, the opposite occurs for low values of $v$. In this case, the system is most of the time in state $B$ since blockages last long. Hence, increasing the time that the link operates in any of the other $N$ states also increases fluctuations.


\subsection{Practical evaluation}

\begin{figure}[t]
	\centering
		\includegraphics{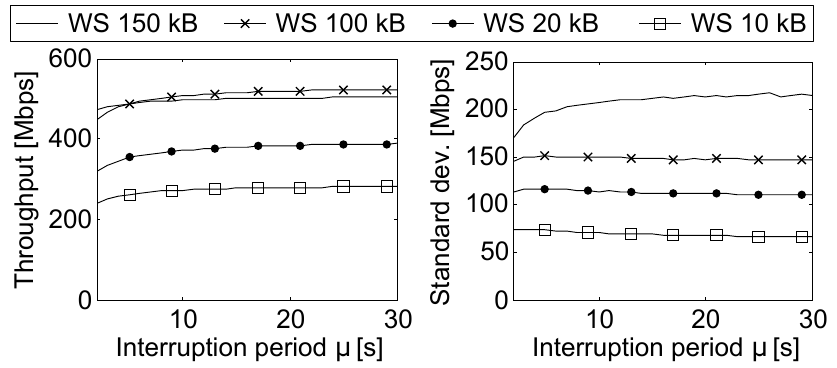}
	\caption{Average throughput for different link traffic loads in practice.}
	\label{fig:res_pract_comb}
\end{figure}

In this section, we present our results based on practical traces collected with our \ac{cots} 60 GHz system as described in Section~\ref{subsec:appl_layer_effects}. We use the traces to compute empirically the number of states $N$ as well as the probabilities $p_H$ and $p_{SLi}$. Further, $p_C = 1$ since we do not observe periodic sector level sweeps in the traces. To control the link load, we adjust the \ac{tcp} \ac{ws}. We configure our Markov model with the above practical parameters to obtain the throughput and the fluctuations for any value of $\mu$. Figure~\ref{fig:res_pract_comb} depicts our result, which is the practical counterpart of our analysis in Figure~\ref{fig:anlyt_load}. We obtain equivalent results in both, which validates our theoretical approach. However, in Figure~\ref{fig:anlyt_load} fluctuations decrease with $\mu$ while in Figure~\ref{fig:res_pract_comb} they are roughly stable. The underlying reason is that in Figure~\ref{fig:anlyt_load} larger $\mu$ values imply that throughput is zero less often, thus decreasing fluctuations. However, in Figure~\ref{fig:res_pract_comb}, iperf provides throughput values at most every $0.5\,\text{s}$. Since blockage often lasts less, we do not observe zero throughput but always the average of the blockage and the following milliseconds. Hence, larger $\mu$ values have a much smaller impact on the dispersion of the throughput values. Further, in Figure~\ref{fig:res_pract_comb} we observe that \ac{tcp} \acp{ws} beyond $100$ kB barely provide additional throughput because aggregation cannot mitigate \ac{mcs} degradations. Still, the respective variance continues to increase significantly beyond that \ac{ws}, i.e., link load. As an example, if we consider $\mu = 5 \text{s}$, the throughput for \ac{ws} $100$ kB is the same as for \ac{ws} $150$ kB but the variance is 41\% larger. This result confirms our hypothesis in Section~\ref{subsubsec:obs_high_load} regarding a link load trade-off.

\section{Conclusion}
\label{sec:conc}

We study the practical impact of transient link blockages on 60 GHz links. We observe that sector level sweeps during a blockage may cause a device to use a suboptimal sector. We analyze two methods to mitigate this. The first one introduces periodic sector level sweeps. To optimize throughput, we find that such sweeps should take place on average as frequently as blockages. However, to minimize throughput variance, sweeps should take place as often as possible. The second method optimizes the link load at the application layer. We find that moderate loads allow frame aggregation to compensate for low \acp{mcs} due to suboptimal sectors. In particular, we show that choosing a suitable link load can reduce throughput variance by $41\%$ compared to a higher load with the same throughput.



\bibliographystyle{abbrv}
\bibliography{biblio}

\begin{thebibliography}{10}

\bibitem{wigig_overview}
C.~Hansen.
\newblock {WiGiG: Multi-Gigabit Wireless Communications in the 60 GHz Band}.
\newblock {\em IEEE Wireless Communications}, 18(6), 2011.

\bibitem{ad_standard}
{IEEE 802.11 Working Group}.
\newblock {IEEE 802.11ad, Amendment 3: Enhancements for Very High Throughput in
  the 60 GHz Band}.
\newblock 2012.

\bibitem{opt_beam_sweep}
B.~Li, Z.~Zhou, H.~Zhang, and A.~Nallanathan.
\newblock {Efficient Beamforming Training for 60-GHz Millimeter-Wave
  Communications: A Novel Numerical Optimization Framework}.
\newblock {\em IEEE Trans. on Vehicular Technology}, 63(2), 2014.

\bibitem{measurements_blockage}
N.~Moraitis and P.~Constantinou.
\newblock {Indoor Channel Measurements and Characterization at 60 GHz for
  Wireless Local Area Network Applications}.
\newblock {\em IEEE Trans. Antennas and Propagation}, 52(12), 2004.

\bibitem{bbs}
T.~Nitsche, A.~Flores, E.~Knightly, and J.~Widmer.
\newblock Steering with eyes closed: Mm-wave beam steering without in-band
  measurement.
\newblock In {\em Proc. IEEE INFOCOM 2015}, 2015.

\bibitem{mobiwac}
A.~Patra, L.~Simi\'{c}, and P.~M\"{a}h\"{o}nen.
\newblock {Smart mm-Wave Beam Steering Algorithm for Fast Link Re-Establishment
  Under Node Mobility in 60 GHz Indoor WLANs}.
\newblock In {\em Proc. of the 13th ACM International Symposium on Mobility
  Management and Wireless Access}, 2015.

\bibitem{challenges60Ghz}
D.~R.C. and R.~Heath~Jr.
\newblock {60 GHz Wireless Communications: Emerging Requirements and Design
  Recommendations}.
\newblock {\em IEEE Vehicular Technology Magazine}, 2(3), 2007.

\bibitem{aggregation_overview}
D.~Skordoulis, Q.~Ni, H.~h.~Chen, A.~P. Stephens, C.~Liu, and A.~Jamalipour.
\newblock {IEEE 802.11n MAC Frame Aggregation Mechanisms for Next-Generation
  High-Throughput WLANs}.
\newblock {\em IEEE Wireless Communications}, 15(1), 2008.

\bibitem{wimi}
S.~Sur, V.~Venkateswarana, X.~Zhang, and P.~Ramanathan.
\newblock {60 GHz Indoor Networking through Flexible Beams: A Link-Level
  Profiling}.
\newblock In {\em Proc. of ACM SIGMETRICS'15}, 2015.

\bibitem{beamscope}
S.~Sur and X.~Zhang.
\newblock {BeamScope: Scoping Environment for Robust 60 GHz Link Deployment}.
\newblock In {\em Proc. of Information Theory and Application Workshop}, 2016.

\bibitem{mtrack}
T.~Wei and X.~Zhang.
\newblock {mTrack: High-Precision Passive Tracking Using Millimeter Wave
  Radios}.
\newblock In {\em Proc. of ACM Mobicom}, 2015.

\bibitem{Zhu2014Mobicom}
Y.~Zhu, Z.~Zhang, Z.~Marzi, C.~Nelson, U.~Madhow, B.~Y. Zhao, and H.~Zheng.
\newblock {Demystifying 60GHz Outdoor Picocells}.
\newblock In {\em Proc. ACM Mobicom}, 2014.

\end{thebibliography}

\end{document}